\documentclass[final]{aastex631}
\usepackage{amsmath,amssymb}
\usepackage{amsmath}
\usepackage{pifont}

\begin{document}

\title{The self-generation of core fields and electron scattering in flux ropes during magnetic reconnection}

\author[0000-0002-0786-7307]{Hanqing Ma}
\affiliation{Department of Physics, University of Maryland, College Park, MD 20740, USA}
\author[0000-0002-9150-1841]{J. F. Drake}
\affiliation{Department of Physics, University of Maryland, College Park, MD 20740, USA}
\author[0000-0002-5435-3544]{M. Swisdak}
\affiliation{IREAP,
University of Maryland,
College Park, MD 20742, USA}

\begin{abstract}
Two-dimensional particle-in-cell simulations with a realistic mass ratio reveal the generation mechanisms of the out-of-plane magnetic field in magnetic islands/flux ropes during magnetic reconnection. In the absence of an initial guide field, reconnection produces a large electron temperature anisotropy ($\sim$4.5) inside magnetic islands that drives the Weibel instability. Strong out-of-plane magnetic fields ($B_z/B_0 \sim 0.4$,  greatly exceeding the Hall field) with a regular bipolar structure grow inside islands. A space-time analysis reveals a one-to-one correspondence between the temperature anisotropy and the development of the Weibel magnetic field. The instability relaxes the anisotropy, but island merging leads to anisotropy reemergence and re-excitation. 
In the presence of a strong ambient guide field ($B_g/B_0 = 0.5$), the electron outflow from the X-point deflects along the separatrices and forms a circular current loop wrapping the flux ropes. 
This flux-rope separatrix current generates an out-of-plane magnetic field that reinforces the ambient guide field, reaching $B_z/B_0 \sim 1.4$. The current can, in some cases, drive the electron Kelvin--Helmholtz instability, which produces electron vortices and strengthens the magnetic field. 
Mergers significantly broaden the islands and further strengthen the field. These self-generated out-of-plane magnetic fields scatter electrons and reduce their temperature anisotropy, which can potentially affect electron heating via Fermi reflection. The simulation results are supported by spacecraft observations suggesting that ambient guide fields can be enhanced within flux ropes in Earth's magnetotail. 
\end{abstract}


\section{Introduction} \label{sec:intro}

Magnetic reconnection is a fundamental plasma process that plays a central role in a wide range of space and astrophysical environments, including solar flares \citep{forbes1991magnetic, shibata1998evidence}, Earth’s magnetotail \citep{nagai2001geotail, bhattacharjee2004impulsive}, and the solar wind \citep{gosling2012magnetic,zank2014particle}. Through reconnection, magnetic energy is efficiently converted into particle kinetic energy, thermal energy, and bulk flows, accompanied by topological changes of magnetic field lines \citep{dahlin2014mechanisms, arnold2021electron, oka2023particle}. A key aspect of reconnection physics is the behavior of electrons in the diffusion region and exhaust, where electron energization produces strong temperature anisotropy and kinetic-scale turbulence \citep{dahlin2014mechanisms, eastwood2009observations, oka2023particle, comisso2024concurrent}. Understanding the mechanisms that regulate electron scattering in these regions is therefore essential for developing accurate models of energy conversion and particle acceleration in collisionless plasmas.

Simulations suggest that electron acceleration during magnetic reconnection is dominated by Fermi reflection in which particles gain energy through repeated reflections between contracting magnetic field lines \citep{drake2006electron, dahlin2014mechanisms, dahlin2016parallel}. In this process, the energy gain is dominated by the parallel motion and scales with the parallel kinetic energy, i.e., $\Delta W \propto v_\parallel^2$ \citep{dahlin2014mechanisms,arnold2021electron}, such that heating occurs primarily along the magnetic field. As a consequence, reconnection-driven Fermi acceleration naturally generates strong temperature anisotropy with $T_\parallel > T_\perp$. Since a sufficiently large anisotropy can drive the firehose instability and stop reconnection, efficient scattering processes that transfer energy from the parallel to the perpendicular direction are therefore crucial for regulating electron heating, relaxing anisotropy, and sustaining continued energization.

A number of mechanisms have been proposed to play a role in electron scattering during magnetic reconnection. Pitch-angle scattering associated with wave--particle interactions driven by kinetic instabilities is commonly discussed \citep{khotyaintsev2019collisionless}. A wide variety of waves and instabilities can be excited in different reconnection regions, and a comprehensive review of these processes is given by \citet{graham2025role}. Among anisotropy-driven instabilities, the Weibel instability \citep{weibel1959spontaneously} is often invoked in space plasmas, yet its role in magnetic reconnection has received relatively limited attention. The Weibel instability generates a non-oscillatory ($\omega \sim 0$) transverse magnetic field with a wave vector predominantly perpendicular to the direction of temperature anisotropy ($k_\perp \gg k_\parallel$ for $T_\parallel \gg T_\perp$). It is widely regarded as a key mechanism for magnetic field generation in space plasmas \citep{schlickeiser2003cosmological, huntington2015observation} and can strongly influence particle dynamics. For example, in collisionless shocks, Weibel-generated magnetic fields scatter particles and enable repeated crossings of the shock front, facilitating particle acceleration through the Fermi mechanism \citep{spitkovsky2008particle}.

Particle-in-cell simulations have demonstrated that reconnection outflows naturally develop strong electron temperature anisotropies that can excite the Weibel instability \citep{lu2011weibel}. However, it has also been suggested that even a weak guide field ($\sim 0.1\,B_0$) could suppress the instability \citep{lu2011weibel}. Alternative mechanisms for Weibel-induced magnetic-field generation associated with electron inflow into the current sheet have been proposed \citep{baumjohann2010magnetic}. In addition, studies of electron--positron (pair) reconnection have shown that the Weibel instability can grow robustly and play an important role in maintaining fast reconnection by broadening the current layer \citep{liu2009weibel, zenitani2008role}. Together, these results suggest that the Weibel instability can be a crucial element of current-layer dynamics in magnetic reconnection, although its effectiveness and dominance may depend sensitively on plasma conditions such as the guide-field strength.

Magnetic reconnection can also generate strong electron shear flows in the outflow region of the current layer that are unstable to the electron Kelvin--Helmholtz (K--H) instability \citep{pu1983kelvin, masson2018kelvin}. The K--H instability produces vortical electron current structures and intense out-of-plane magnetic fields that can efficiently scatter electrons. Simulations have shown that the K--H instability can generate multiple vortex structures, which play an important role in driving secondary magnetic reconnection and enhancing turbulence in the reconnection outflow \citep{fermo2012secondary, huang2015magnetic, zhong2018evidence}. Most of these simulations are carried out in the presence of a guide field, and the relative importance of the Weibel instability versus shear-driven instabilities in realistic reconnection configurations therefore remains uncertain.

Observations, on the other hand, frequently report strong out-of-plane magnetic fields near the centers of magnetic islands during reconnection events, often accompanied by enhanced fluxes of energetic electrons \citep{chen2008observation,chen2009multispacecraft,wang2010observations, wang2010situ}. In the magnetotail, measurements commonly reveal a significant out-of-plane magnetic-field component within flux ropes (e.g., \citep{huang2016mms,akhavan2018mms,sun2019mms,zhou2021observations, liu2025mms}). The physical origin of these strong core magnetic fields remains an open question. The formation of a core field inside magnetic islands is often attributed to the compression of the initial guide field during island contraction and merging associated with the loss of internal plasma \citep{ma1992enhancements, ma1994core, hesse1996simple, jin2004core, drake2006formation}. \citet{scholer1988strong} interpreted enhanced core fields in the framework of a bursty reconnection model, while \citet{karimabadi1999magnetic} carried out hybrid reconnection simulations and attributed the large core fields to Hall-generated currents. In this context, anisotropy- or shear-flow-driven instabilities may provide additional or complementary mechanisms for generating and sustaining strong core magnetic fields inside magnetic islands.

In this work, we use two-dimensional (2D) particle-in-cell (PIC) simulations of magnetic reconnection with a realistic proton-to-electron mass ratio to investigate the development of flux-rope core magnetic fields driven by the Weibel instability, flux-rope separatrix currents, and the electron Kelvin--Helmholtz instability.
We find that the Weibel instability is the dominant mechanism responsible for generating strong out-of-plane magnetic fields in the absence of a guide field. Even in the presence of a strong guide field ($B_z/B_0 \sim 0.5$), pronounced self-generated out-of-plane magnetic fields develop inside magnetic islands. With the guide field, the flux-rope separatrix current and shear K--H instability are responsible for the most intense magnetic structures. Island merging plays a key role in maintaining and further amplifying the core magnetic field strength. Relevant observational evidence is also discussed.

The remainder of the paper is organized as follows. In Sec.~II, we describe the 2D PIC simulation model. In Sec.~III, we present the simulation results, first for the case without an ambient initial guide field and then for the case with a strong initial guide field. Section~IV contains a discussion of the results and a summary of the main conclusions.

\section{Simulation setup} \label{sec:simeth}
We perform two-dimensional particle-in-cell (PIC) simulations with a realistic mass ratio using the code {\tt p3d} \citep{zeiler2002three}, which advances particle trajectories by integrating the relativistic Newton--Lorentz equations and evolves the electromagnetic fields using Maxwell’s equations. The proton-to-electron mass ratio is $m_i/m_e = 1836$, with the term ``ion'' in this work referring exclusively to protons. The simulation domain contains two Harris-type current sheets, with the initial magnetic field specified as $B_x(y) = B_0 \left[ \tanh\!\big((y - 0.25\,l_y)/w_0\big) 
- \tanh\!\big((y - 0.75\,l_y)/w_0\big) - 1 \right]$, where $w_0$ is the half-width of the current sheet. We choose $w_0$ = $0.012d_i\approx0.5d_e$, where $d_i = c/\omega_{pi}$ is the ion inertial length and $\omega_{pi}={(4\pi n_ie^2/m_i)}^{1/2}$ is the ion plasma frequency. 
We first carry out a simulation with no initial guide field and subsequently introduce an initial guide field, $B_z=B_g$, to examine its influence on the system. Both electrons and ions are initialized with Maxwellian velocity distributions. To maintain an equilibrium configuration, the density profile is given by $n_e = n_i = [\,\mathrm{sech}^2((y-0.25\,l_y)/w_0) + \mathrm{sech}^2((y-0.75\,l_y)/w_0)\,]/(2T) + n_0$. The system size is $l_x = 2l_y=15.36d_i$ and periodic boundary conditions are applied in both directions. The simulation time step is $dt = 10^{-5} \Omega_i^{-1}$, $v_{Ai} = B_0/\sqrt{4\pi n_0m_i} \approx 0.01c$, and $\beta = 8\pi n_0T/B_0^2 \approx 0.05$. The simulation uses 500 particles per species per cell and employs a grid of 6144 by 3072 cells. 


\section{Simulation Results}

\subsection{Zero Guide Field Case}

\begin{figure}[ht!]
\centering
\includegraphics[scale=0.8]{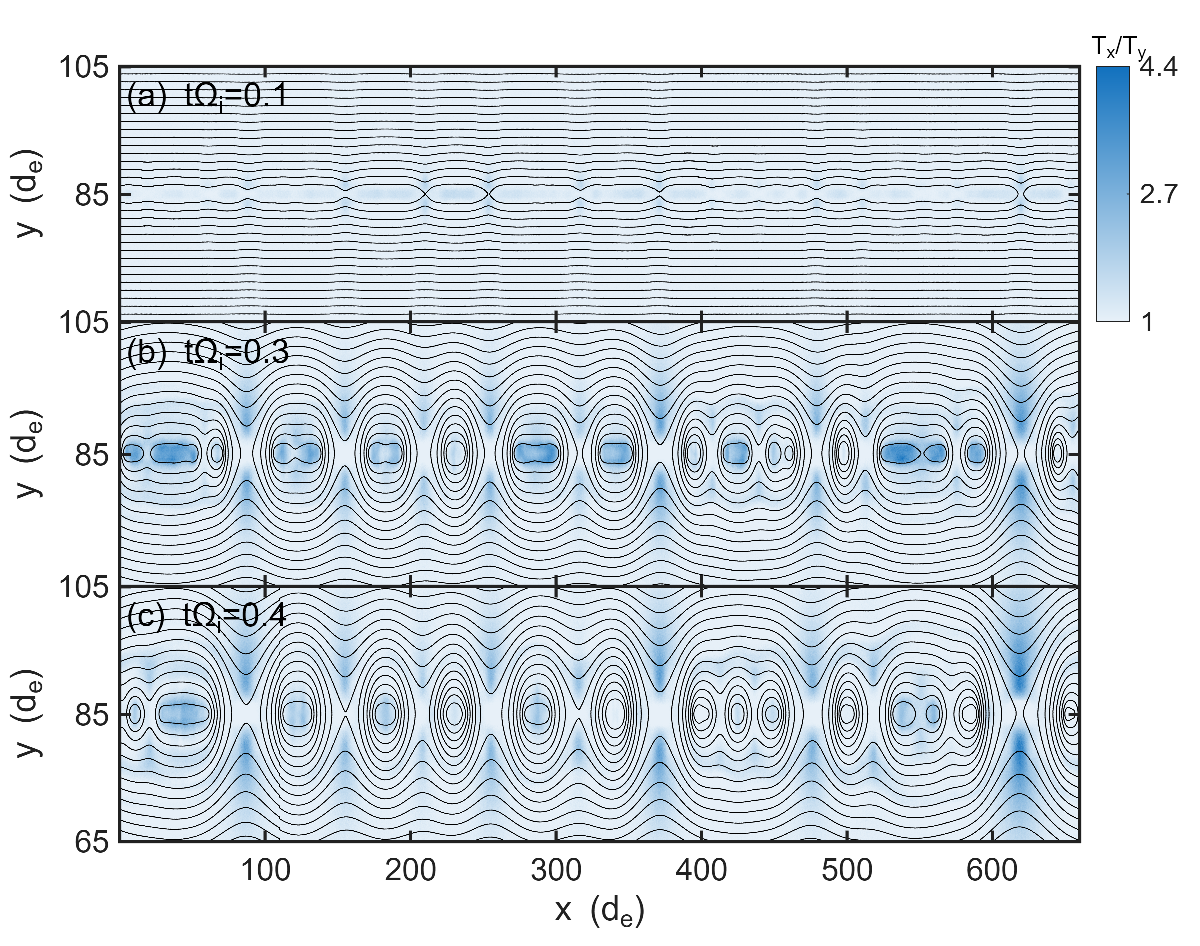}
\caption{Structure of the lower current sheet (the simulation domain contains two current sheets) at (a) $t\Omega_i =0.1$, (b) $t\Omega_i=0.3$ and (c) $t\Omega_i=0.4$. Black lines denote magnetic field lines and the color scale shows the electron temperature anisotropy $T_x/T_y$.}
\label{sheet0}
\end{figure}

Figure~\ref{sheet0} shows the dynamics in the lower current sheet. In panel (a), anti-parallel magnetic field lines break and reconnect, forming multiple X-lines, with magnetic islands developing between them. The increase in temperature anisotropy upstream of the X-lines arises from the conservation of the magnetic moment $\mu$. In panel (b), electrons are accelerated at the X-lines, and the resulting outflows produce strong temperature anisotropies inside the islands, reaching $T_x/T_y \sim 4.5$. This strong anisotropy triggers the Weibel instability within the region where $B_x$ is small. In panel (c), the anisotropy inside the islands is substantially reduced due to the scattering from the instability, with $T_x/T_y \sim 1$. Compared with panel (b), the islands also broaden in the $y$ direction.

\begin{figure}[ht!]
\centering
\includegraphics[scale=0.8]{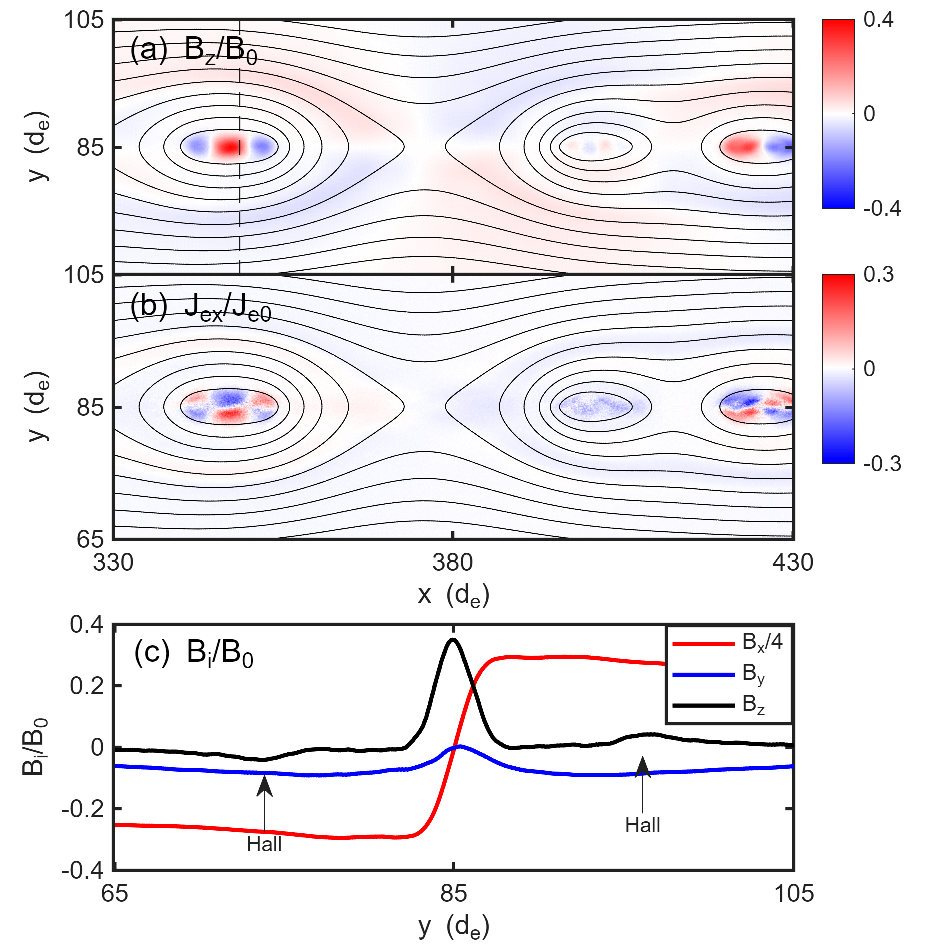}
\caption{(a) Out-of-plane magnetic field $B_z$ at $t\Omega_i \approx 0.37$. Black contours indicate magnetic field lines. (b) The $x$-direction electron current $J_{ex}$ at the same time. The electron current is normalized by $J_{e0}=en_0v_{Ae}$ where $v_{Ae}$ is the electron Alfv$\acute{e}$n speed. (c) Magnetic field components along the vertical dashed line in (a).  The perturbations due to the usual Hall field are indicated.  }
\label{Bz2d0}
\end{figure}

Figure~\ref{Bz2d0} shows a zoomed-in view around a reconnection site. In panel (a), the Weibel instability at the island center generates strong out-of-plane magnetic fields with amplitudes reaching $B_z/B_0 \sim 0.4$, greatly exceeding the quadrupolar Hall magnetic field. The resulting $B_z$ alternates in sign along the $x$ direction, while only a single wavelength develops in the $y$ direction. In principle, linear theory predicts that the wave vector of the Weibel instability is oriented along the $y$ direction, i.e., perpendicular to the direction of temperature anisotropy, leading to the formation of multiple magnetic filaments \citep{weibel1959spontaneously, morse1969numerical, stockem2009pic}. In the present simulation, however, the growth of Weibel modes in the $y$ direction is constrained by the narrow width of the current sheet ($\sim 0.5\,d_e$), which is bounded on both sides by regions of strong $B_x$, the ambient magnetic field. For the same reason, no clear Weibel signatures are observed upstream of the X-line, even in the presence of strong temperature anisotropy. 

Figure~\ref{Bz2d0}(b) shows enhanced electron fluxes inside the magnetic island, where the electron current exhibits a crossover pattern. This current structure is consistent with the spatial variation of the Weibel-generated magnetic fields along the $x$ direction. These bipolar out-of-plane magnetic structures therefore result from the Weibel instability rather than from compression of the ambient magnetic field. In contrast, the ion flow (not shown), with $J_{ix}/J_{e0}$ ranging from $-0.08$ to $0.08$, does not participate in the instability.

Figure~\ref{Bz2d0}(c) shows the variations of the magnetic-field components along a vertical cut (indicated by the dashed line in panel (a)) through the magnetic island. The out-of-plane magnetic field $B_z$ exhibits a pronounced peak at the island center, where $B_x$ reverses sign. In contrast, the Hall magnetic field appears in regions displaced from the island center, with an amplitude that is significantly smaller than that of the Weibel-generated $B_z$. It is worth noting that, for a horizontal cut through the island, $B_z$ exhibits multiple bipolar peaks at locations where $B_y$ reverses sign. Similar signatures have been reported in magnetotail observations. For example, Cluster measurements of the  structure of a magnetic island reveal a peak in the cross-tail magnetic field in the island core and a peak in the total magnetic-field magnitude at the same location \citep{wang2010observations}. There was no ambient guide field in these observations so the large guide field did not result from the compression of an ambient guide field. Further discussion of this event is presented later, following an exploration of the role of the guide field.

\begin{figure}[ht!]
\centering
\includegraphics[scale=0.8]{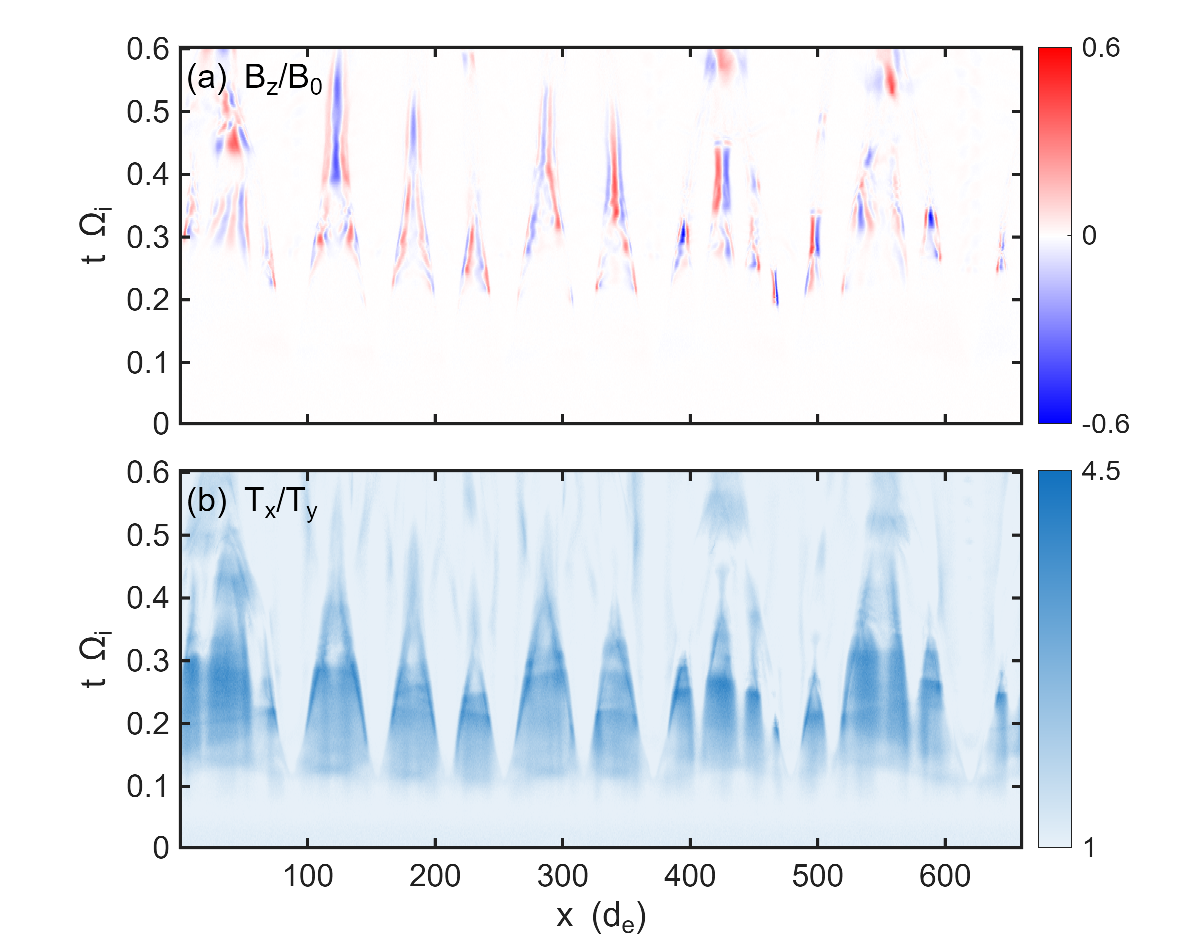}
\caption{Space--time plot of the out-of-plane magnetic field $B_z$ and electron temperature anisotropy $T_x/T_y$ in the center of the lower current sheet.} 
\label{spacetimeplot0}
\end{figure}

Figure~\ref{spacetimeplot0} shows the time evolution of the out-of-plane magnetic field $B_z$ and the electron temperature anisotropy $T_x/T_y$ at the center of the lower current sheet. The anisotropy begins to develop at $t\Omega_i \approx 0.15$ as electrons are accelerated by reconnection. Shortly thereafter, at $t\Omega_i \approx 0.2$, an enhancement of $B_z$ emerges as the Weibel instability is triggered by the growing anisotropy. The clear one-to-one temporal correspondence between the anisotropy and the growth of $B_z$ confirms that the Weibel instability is responsible for generating the strong out-of-plane magnetic field in the island center.
The anisotropy starts to decrease at $t\Omega_i \approx 0.3$ as a result of pitch-angle scattering by the Weibel-generated fields. As the anisotropy is reduced and the instability drive weakens, the $B_z$ signal decays. A magnetic-island merging event occurs near $d_e \approx 450$ at $t\Omega_i \approx 0.3$. Following this merger, the anisotropy reappears with a reduced amplitude at $t\Omega_i \approx 0.5$, accompanied by a renewed increase in $B_z$ around $t\Omega_i \approx 0.6$. This sequence demonstrates that island coalescence regenerates temperature anisotropy, which again drives the growth of the Weibel instability and the associated generation of the out-of-plane magnetic-field.

\begin{figure}[ht!]
\centering
\includegraphics[scale=0.8]{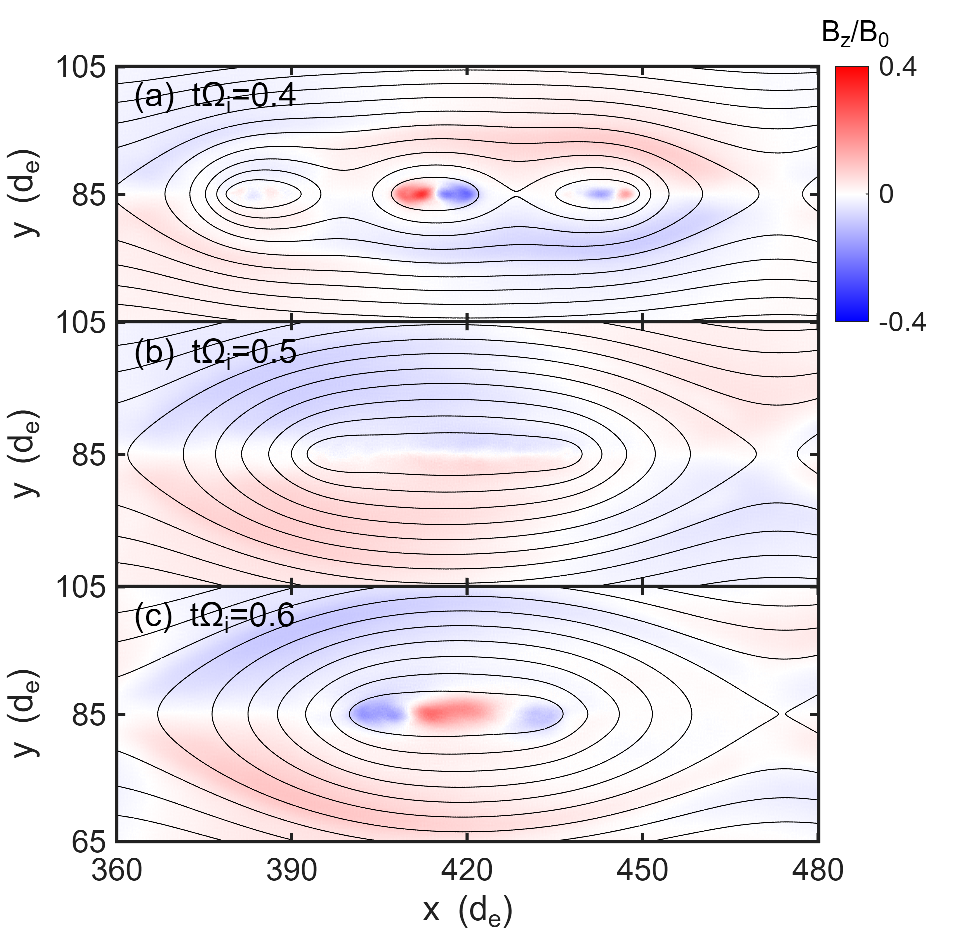}
\caption{Weibel evolution during magnetic island merger. Black contours denote magnetic field lines, and the color scale shows the out-of-plane magnetic field $B_z/B_0$.} 
\label{islandmerge0}
\end{figure}

Figure~\ref{islandmerge0} shows the impact of island on Weibel generated fields. The three smaller islands in panel (a) coalesce into a larger island in panel (b). During this process, the initial $B_z$ structures shown in panel (a) dissipate and partially cancel. Island merging leads to the re-emergence of temperature anisotropy as magnetic field lines contract in the $x$ direction \citep{drake2006formation,drake2012power,zhou2014plasma,che2019brief}. The regenerated anisotropy excites the Weibel instability again, as shown in panel (c), although the resulting $B_z$ amplitude is weaker than that produced during the initial reconnection phase in panel (a). This merging process highlights that island dynamics play a critical role in regenerating anisotropy and sustaining Weibel activity well beyond the onset of reconnection. Thus, strong out-of-plane magnetic-field signatures may persist longer than expected as a result of island merging as the release of magnetic energy continues.

Since the strength of the initial guide field is expected to play an important role, we have varied this parameter.  For weaker guide fields ($B_g/B_0 \sim 0.2$), the peak anisotropy is reduced because the guide field suppresses reconnection-driven electron acceleration \citep{huba2005hall, wan2008electron, dahlin2015electron}. The bipolar magnetic-field structure produced by the Weibel instability remains significant, but is shifted by the presence of the guide field: both the minimum and maximum values of the out-of-plane magnetic field are offset by the guide field, while their difference remains approximately unchanged.  The next section presents simulation results with a uniform strong guide field ($B_g/B_0 = 0.5$) applied throughout the entire simulation domain.

\subsection{Finite Guide Field Case}

\begin{figure}[ht!]
\centering
\includegraphics[scale=0.8]{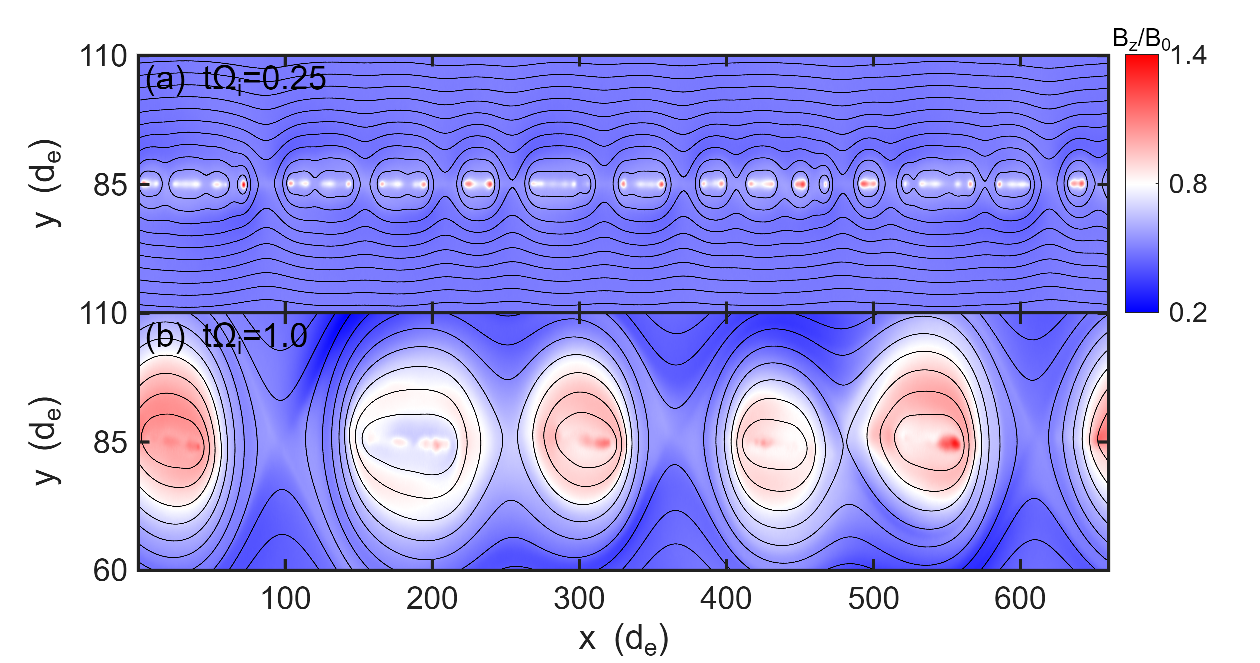}
\caption{The lower current sheet at (a) $t\Omega_i = 0.25$ and (b) $t\Omega_i = 1$ in a simulation with an initial guide field $B_g/B_0 = 0.5$. Black contours denote magnetic field lines, and the color scale shows the out-of-plane magnetic field $B_z/B_0$.}
\label{Bz2dguide05}
\end{figure}

Figure~\ref{Bz2dguide05} shows the structure of the lower current sheet in the simulation with an initial guide field $B_g/B_0 = 0.5$. During the early phase of reconnection, shown in panel (a), enhanced out-of-plane magnetic fields appear inside the magnetic islands. This enhancement is strongest near the ends of the island, with amplitudes reaching $B_z/B_0 \sim 1$. These $B_z$ enhancements are unidirectional and aligned with the initial guide field. The enhanced out-of-plane magnetic-field structures are generated by the flux-rope separatrix currents and electron Kelvin--Helmholtz instability.

At the end of the simulation, shown in panel (b), island merging produces larger magnetic islands with sizes of $\sim 2\,d_i$ in the $x$ direction. The islands are also significantly broadened in the $y$ direction, extending well beyond the initial current-sheet width. The resulting islands exhibit a two-layer structure in $B_z$, with a localized peak field of $B_z/B_0 \sim 1.4$ at the center and a broader outer region of weaker fields, still exceeding the ambient guide-field strength, with $B_z/B_0 \sim 0.8$. The mechanism that produces these flux rope core fields is discussed later.

\begin{figure}[ht!]
\centering
\includegraphics[scale=0.8]{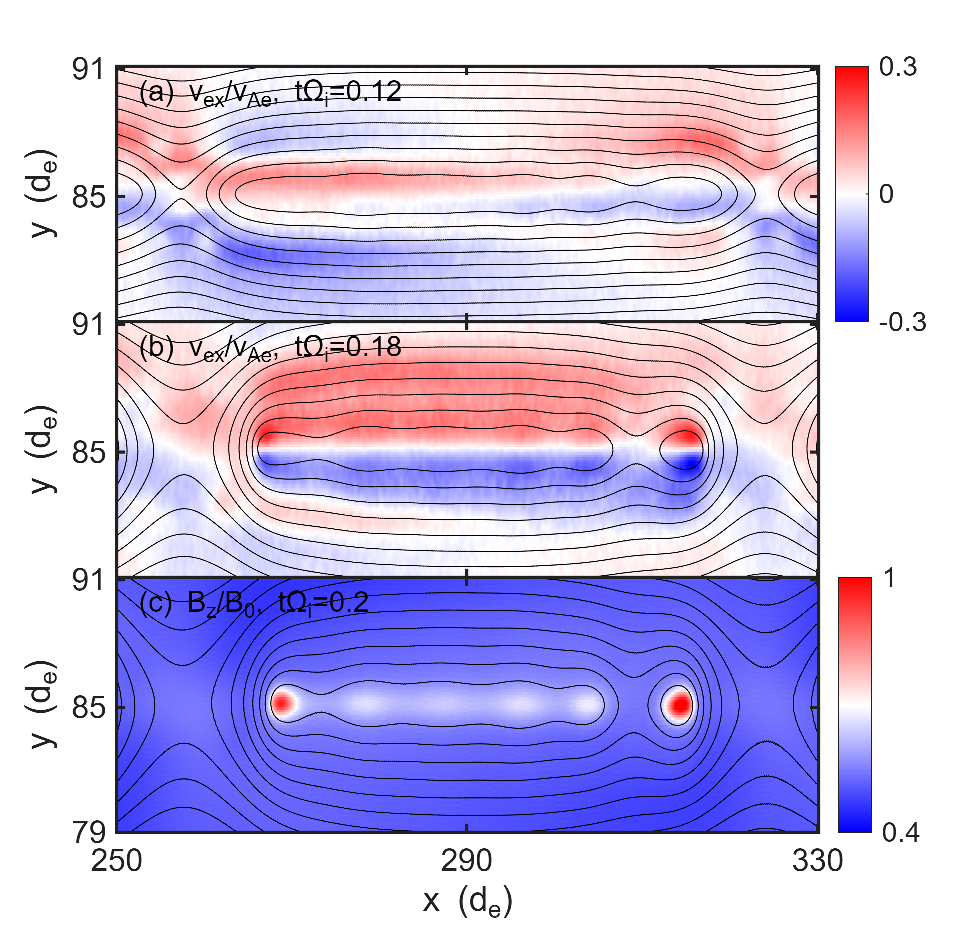}
\caption{Electron bulk flows in the $x$ direction, $v_{ex}/v_{Ae}$, between two X-lines at (a) $t\Omega_i = 0.12$ and (b) $t\Omega_i = 0.18$. (c) The out-of-plane magnetic field $B_z/B_0$ distribution at $t\Omega_i = 0.2$.}
\label{shearandcurrent}
\end{figure}

Figure~\ref{shearandcurrent} shows the electron bulk velocity $v_{ex}$ and out-of-plane magnetic field in a representative magnetic island. During the early phase of reconnection, shown in panel (a), electrons originating from the left X-line move in the positive $x$ direction along the upper separatrix, while those from the right X-line move in the negative $x$ direction along the lower separatrix. This deflection produces a velocity shear within the magnetic island, with flow speeds comparable to $v_{Ae}$.
As shown in panel (b), the electrons from the adjacent x-lines combine to form a closed current loop. Near the island center, the velocity shear triggers the electron Kelvin--Helmholtz instability, which fragments the initial current layer. Two dominant current loops form near the two ends of the island, while several weaker circular currents develop within the island interior. All current loops rotate counterclockwise, generating out-of-plane magnetic fields aligned with the initial guide field. Panel (c) shows the intense $B_z$ generated by these current loops, with the strongest peaks located near the two X-lines and weaker peaks appearing near the island center. 

Figure~\ref{shearandcurrent} corresponds to a case in which the two X-lines are sufficiently separated, allowing the Kelvin--Helmholtz instability to fully develop. If the two X-lines are sufficiently close to each other, as in the case shown in Figure~\ref{Bz2dguide05}(a) at $x \sim 500\,d_e$, only a single strong $B_z$ peak is formed. It is worth noting that the circular currents generate an out-of-plane magnetic field that enhances the initial guide field at the island center, whereas outside the island the magnetic field associated with these currents is opposite to the guide field, leading to a weaker surrounding guide field so that the net magnetic flux is conserved. These enhanced magnetic-field structures arise from counter-streaming, sheared electron flows and the electron Kelvin--Helmholtz instability, rather than from compression of the ambient guide field.

\begin{figure}[ht!]
\centering
\includegraphics[scale=0.2]{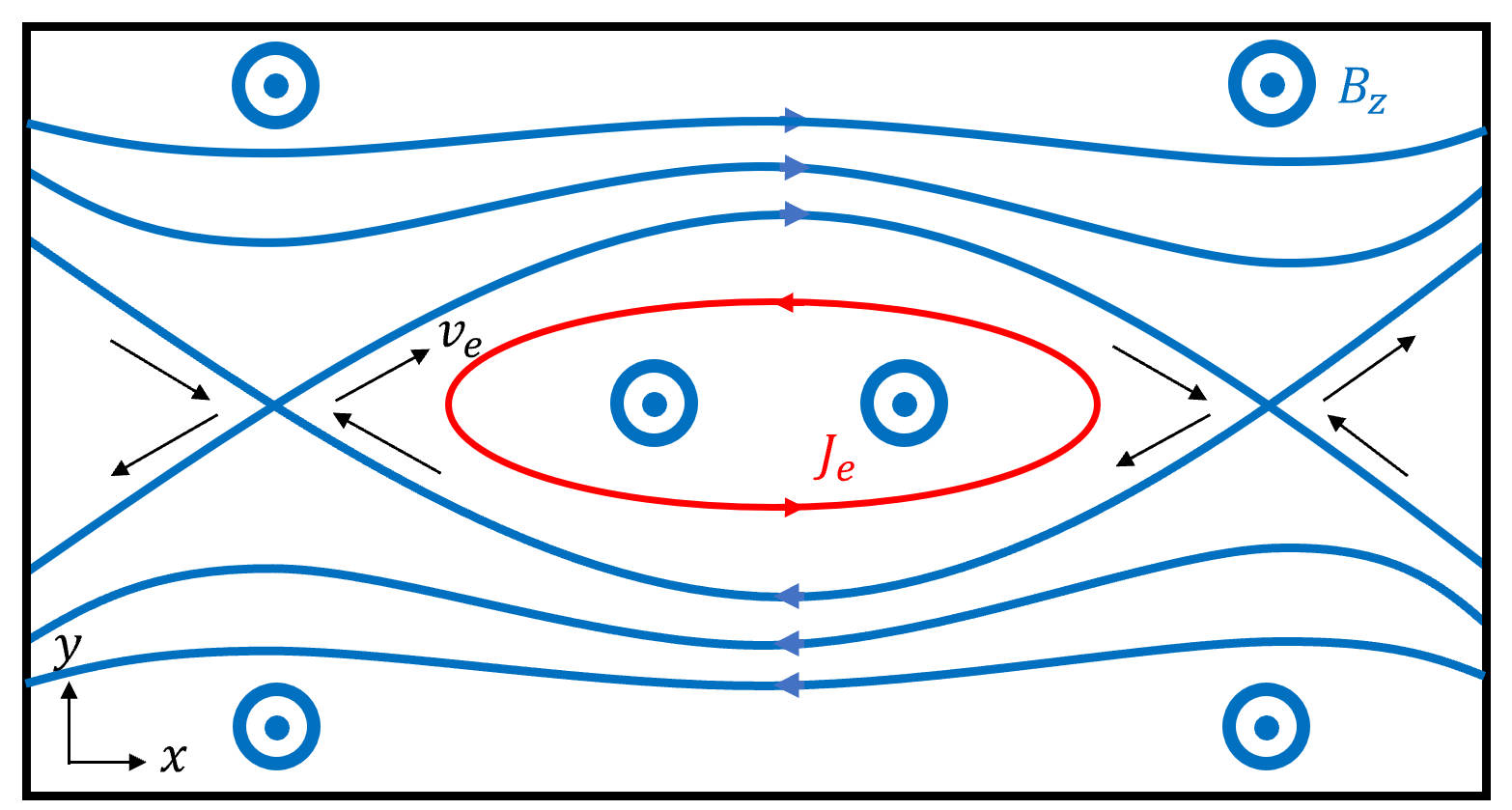}
\caption{Cartoon diagram illustrating how the flux-rope separatrix current enhances the core magnetic field. Blue lines denote magnetic field lines, and the out-of-plane magnetic field $B_z$ is represented by blue circles. Black arrows indicate electron velocities, and red arrows denote the electron current.}
\label{diagram}
\end{figure}

In Figure~\ref{diagram} we present a schematic illustrating the electron motion within the flux rope and the associated enhancement of the core magnetic field. The current around the x-line flows into the plane and hence the electrons flow out of the plane. Near the left x-line, and just inside of the flux rope, the magnetic field points out-of-the-plane and towards the x-line near the lower separatrix and out-of-the-plane and away from the x-line near the upper separatrix. Thus, the electrons within the flux rope flow towards the x-line inside the lower separatrix and away from the x-line inside the upper separatrix. These flows near the left x-line combine with corresponding flows at the right x-line to form a closed current loop inside the island.  We refer to this current loop as the flux-rope separatrix current. The resulting current generates an out-of-plane magnetic field aligned with the guide field, thereby enhancing the magnetic field in the island core. The enhancement in the guide field is boosted in the present simulations because of the realistic mass ratio. The electron separatrix flows are comparable in magnitude to the electron Alfv\'en speed,  as can be seen in Figure~\ref{shearandcurrent}. The time separation between Figures~\ref{shearandcurrent}(a) and (b) is only $0.06\Omega_i^{-1}$. Even during this short time interval the electron currents localized around the two adjacent x-lines were able to combine into a single current loop within the flux rope.

\begin{figure}[ht!]
\centering
\includegraphics[scale=0.8]{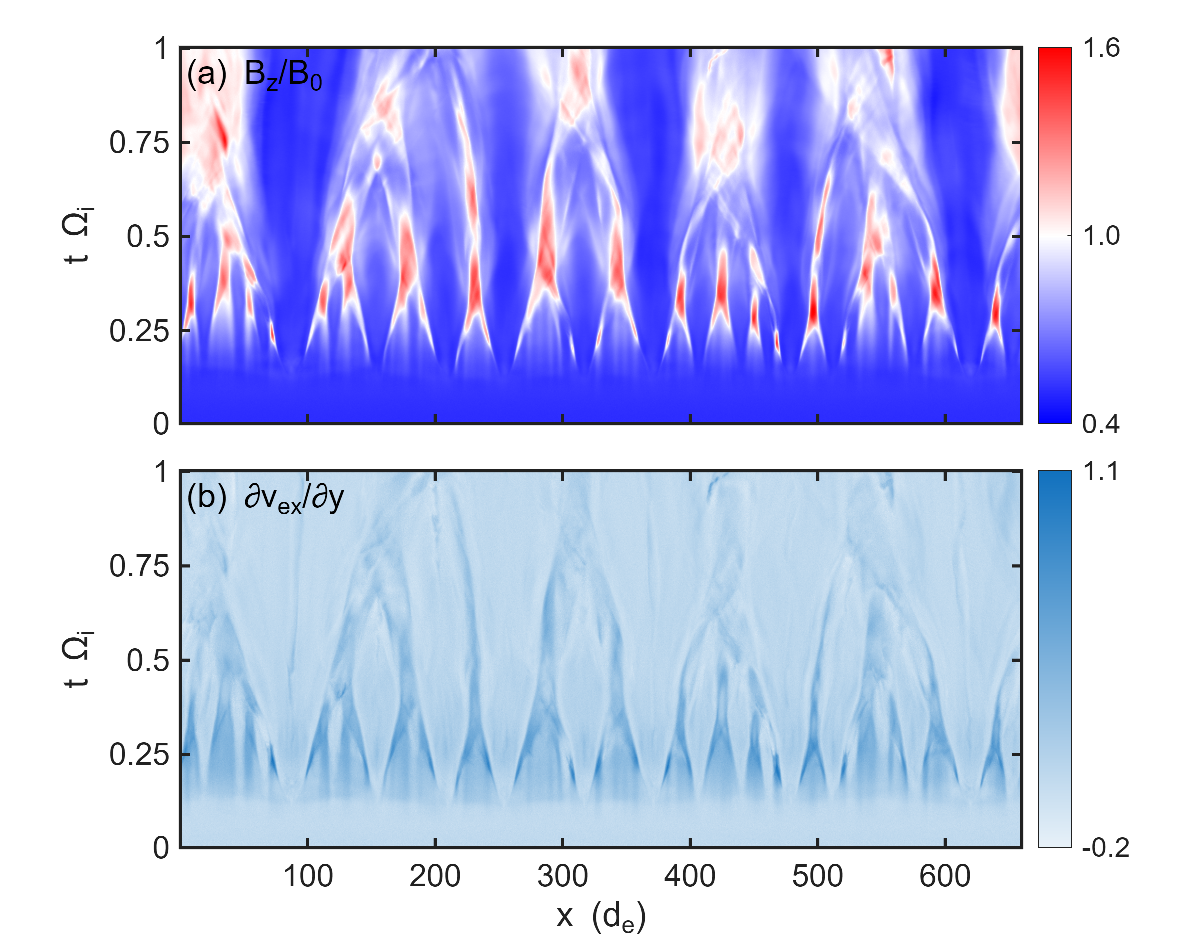}
\caption{Space–time plot of the out-of-plane magnetic field $B_z/B_0$ and $\partial v_{ex}/\partial y$ at the center of the lower current sheet with guide field $B_g/B_0=0.5$. The quantity $\partial v_{ex}/\partial y$ is normalized by $v_{Ae}/d_e$ and is a measure of the sheared electron flow in the core of the islands. } 
\label{spacetimeguide05}
\end{figure}

Figure~\ref{spacetimeguide05} shows the space--time evolution of the out-of-plane magnetic field $B_z$ and the electron current gradient $\partial v_{ex}/\partial y$ at the center of the lower current sheet. The quantity $\partial v_{ex}/\partial y$ measures the velocity shear across the current sheet and is calculated by averaging the slope near the current-sheet center. A clear one-to-one correspondence between $B_z$ and the current gradient is observed, indicating that the enhancements of the out-of-plane magnetic field are produced by the electron velocity shears. The electron temperature anisotropy (not shown) is reduced in this case because the guide field suppresses reconnection-driven electron energization \citep{huba2005hall, wan2008electron, dahlin2015electron}. Despite the weaker temperature anisotropy, the peak $B_z$ is significantly enhanced compared with the zero-guide-field case, reaching values approximately three times larger than the initial guide-field strength. The space--time plot further indicates that the dominant growth of $B_z$ occurs during island-merging events starting from  $t\Omega_i \approx 0.3$. After $t\Omega_i \approx 0.7$, the out-of-plane magnetic field saturates and remains largely stable at a level of $B_z/B_0 \sim 1$.

\begin{figure}[ht!]
\centering
\includegraphics[scale=0.8]{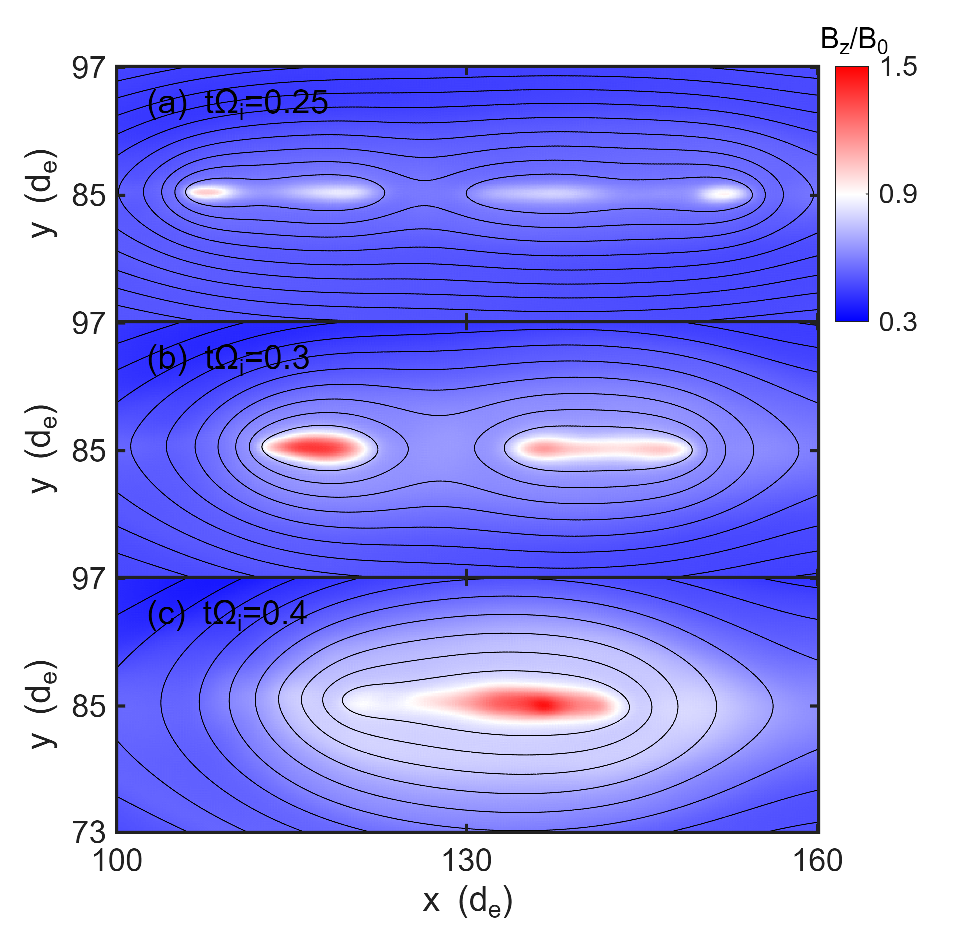}
\caption{Magnetic island merging with an initial guide field $B_g/B_0 = 0.5$. Black contours denote magnetic field lines, and the color scale shows the out-of-plane magnetic field $B_z/B_0$.} 
\label{islandmerge05}
\end{figure}

Figure~\ref{islandmerge05} shows an example of magnetic-island merging in the presence of a guide field. From panel (a) to panel (b), the two islands contract in the $x$ direction, accompanied by a progressive increase in the out-of-plane magnetic field $B_z$. From panel (b) to panel (c), the two islands merge into a single larger island, and the amplitude of $B_z$ reaches values of approximately $1.5\,B_0$. During the merging process, the electron density does not increase along with the enhancement of the out-of-plane magnetic field in the island core. Instead, we observe an increase in $J_{ex}$ inside the island, as the contracting magnetic field lines accelerate electrons in the $x$ direction. In addition, calculations of $\boldsymbol{\nabla} \boldsymbol{\cdot} \mathbf{v}$ do not show a significant compression during island growth and merging. We therefore emphasize that the strong enhancement of $B_z$ observed here is not simply a consequence of compression of flux ropes as suggested in earlier studies \citep{ma1992enhancements, ma1994core, hesse1996simple, jin2004core,drake2006formation}, but rather results from the generation and amplification of electron separatrix currents.

\section{Conclusions}

We performed two-dimensional particle-in-cell simulations with a realistic ion--electron mass ratio to investigate the formation of out-of-plane magnetic field inside flux ropes/magnetic islands during reconnection. In the absence of a guide field, strong temperature anisotropy develops in the reconnection exhaust and triggers the Weibel instability, which generates bipolar out-of-plane magnetic fields along the current sheet within the magnetic islands. In contrast, when a strong guide field is present, the electron outflows are deflected by the guide field to form flux-rope separatrix currents, a circular current loop wrapping the flux rope, which generates an out-of-plane magnetic field that reinforces the initial guide field. The extended sheared flow near the center of the flux rope becomes unstable to the electron Kelvin--Helmholtz instability, leading to the formation of localized rings of electron current and associated regions of enhanced magnetic field $B_z$. Island merging further helps to sustain and, in some cases, amplify the core magnetic field. These magnetic-field structures constitute an important element of reconnection dynamics and are expected to play a key role in scattering energetic electrons during reconnection.

The rate of reconnection-driven particle acceleration, which is dominated by Fermi reflection, is proportional  to a particle's velocity parallel to the magnetic field. Consequently, scattering processes that transfer energy from the parallel to the perpendicular direction can modify reconnection-driven energy gain. In our simulations, the instability-generated out-of-plane magnetic fields strongly scatter the electron outflows, leading to a substantial increase in the perpendicular velocity. Such scattering is therefore expected to influence the energy gain during island merging.

Although this paper focuses primarily on two limiting cases, $B_g = 0$ and $B_g/B_0 = 0.5$, we also performed a series of simulations with intermediate guide-field strengths. In all cases, we observed the growth of strong out-of-plane magnetic fields $B_z$ near the centers of magnetic islands. As discussed previously, the dominant mechanism responsible for $B_z$ generation is the Weibel instability for $B_g=0$, whereas the flux-rope separatrix currents and electron Kelvin--Helmholtz instability dominate for $B_g/B_0 = 0.5$. A key difference between these two cases is the direction of the enhanced out-of-plane magnetic field (see Figures~\ref{Bz2d0}(a) and \ref{Bz2dguide05}(a)), which is rooted in the direction of the electron current (see Figures~\ref{Bz2d0}(b) and \ref{shearandcurrent}(b)). The transition between these two regimes is not sharp. For intermediate and weak guide-field strengths (e.g., $B_g/B_0 \sim 0.2$), the guide field acts as a perturbation to the electron outflow, and both mechanisms appear to operate simultaneously. These results suggest that the development of a strong peak in $B_z$ within magnetic islands is a universal feature of magnetic reconnection. Such features provide valuable insight into the structure of electron-scale outflows and are signatures of the electron scattering that occurs when accelerated particles are ejected into growing magnetic islands.

Magnetic flux tubes, or plasmoids, are frequently observed to possess strong core magnetic fields that are comparable to, or even exceed, the strength of the ambient guide field \citep{huang2016mms, akhavan2018mms,sun2019mms,zhou2021observations,liu2025mms}. Traditionally, the enhancement of the core magnetic field has been interpreted as a consequence of flux-tube compression associated with the loss or evacuation of internal plasma \citep{ma1992enhancements,ma1994core,hesse1996simple, jin2004core,drake2006formation}. These interpretations are primarily based on magnetohydrodynamic or hybrid simulations, which do not fully capture electron-scale dynamics or the effects of kinetic micro-instabilities. In contrast, our PIC simulations, enabled by the use of the realistic proton-to-electron mass ratio, explicitly resolve electron dynamics and reveal the importance of electron-scale processes in generating and sustaining strong core magnetic fields. These results provide a new perspective on the origin of enhanced core fields in flux ropes and suggest that kinetic effects may play a more significant role than previously recognized.

One important consideration for observations is that a flux rope can propagate after its formation \citep{moldwin1994observations,kiehas2012formation}. A flux rope may form in a region with a finite guide field and subsequently move into a location with little or no guide field. Therefore, the strength of the ambient guide field outside the flux rope should not necessarily be interpreted as the guide field experienced during its formation. In addition, as shown in Figures~\ref{Bz2d0}(a) and \ref{Bz2dguide05}(a), there are valleys between adjacent magnetic-field peaks. If a spacecraft trajectory passes through such a valley, it may not detect a strong out-of-plane magnetic field even though a significant guide field might be present at another location within the flux rope. These factors should be taken into careful consideration when comparing simulation results with observations.

The initial current sheet width used in these simulations is relatively narrow, on the order of $d_e$. In space plasmas, the current sheet thickness can vary substantially and a width of $\sim d_e$ represents the lower bound inferred from magnetotail observations \citep{phan2018electron,torbert2018electron,zhou2021observations,hubbert2022electron}. However, as shown in Figures~\ref{islandmerge0} and \ref{Bz2dguide05}, the magnetic islands significantly broaden at later times, reaching widths of order $d_i$ for which proton dynamics become important. Even in these larger scale flux ropes the enhancement of the out-of-plane magnetic field remains pronounced. We also tested a case with an initial current sheet width of $2 d_e$, with the result being a delayed reconnection onset and a weaker, but still significant, core field. How the generation of out-of-plane magnetic fields depends on parameters such as the plasma beta and current-sheet asymmetry remains an open question and will be an important topic for future investigations.


\begin{acknowledgements}
The authors were supported by NASA grants 80NSSC22K0352, and 80NSSC20K1813, and NSF grant PHY2109083. The simulations were performed in the National Energy Research Scientific Computing Center (NERSC). 
The simulation data and analysis products used in this study are publicly available on Zenodo at \url{https://doi.org/10.5281/zenodo.18829455}.

\end{acknowledgements}

\bibliographystyle{aasjournal}   
\bibliography{paperbib}

\end{document}